\documentclass[
    ,final            
  ]
  {aipproc}

\layoutstyle{6x9}

\begin{document}

\title{Features of elastic scattering at
small $\bf t$\\ at the LHC
}


\author{O.V. Selyugin}{
  address={ Bogoliubov
 Laboratory of Theoretical Physics, JINR, 141980, Dubna, Russia
        } }

\author{J.-R. Cudell}{
  address={IFPA, Dept. AGO, B\^at. B5a,
Universit\'e de Li\`ege, B4000
  Li\`ege, Belgium} }

\begin{abstract}
 The problems linked with the extraction of the basic parameters of
 the hadron elastic scattering amplitude at the LHC are explored.
It is shown that one should take into account the saturation regime
which will lead to new effects at the LHC.
\end{abstract}

\maketitle
In the nearest future, new experiments will measure the elastic
scattering amplitude at the LHC. We hope that they will improve the
theoretical understanding of this basic
object of hadronic physics.
Now models predict a wide range of possibilities,
for example for the growth of  $\sigma_{tot}(s)$. However, we had such a
situation a long time ago.
\begin{quote}
``With the construction of large accelerators, it is hoped
that the mysteries of high-energy scattering will unfold
in the near future." \\ \hglue 4cm {Hung Cheng, Tsai Tsun Wu, Phys. Rev.
Lett. (1970).}
\end{quote}
Long after, we met the problems linked with  different results for
the size of $\rho$, the ratio of the real part to the imaginary part
of the scattering amplitude [UA4  measured $\rho=0.24$ and UA4/2
 $\rho=0.139$] and then,
at the Tevatron,  $\sigma_{tot}(s)$ is ill-determined
[$\sigma_{tot}= (71.42 \pm 2.41) $ mb, the results of E811 experiment or $\sigma_{tot}= (80.03  \pm
2.24) $ mb according to CDF.]\footnote{Note that it is likely that this
difference
reflects the true errors of
the procedure extracting $\sigma_{tot}$ from the experimental data
through the luminosity-independent method. }

The theoretical predictions for $\sigma_{tot}$ at the LHC
cover a wide range, from $80 \ $mb up to $230 \ $mb.
For elastic scattering, several models predict very different
behaviours, such as, for example, oscillations, or a non-exponential
dependence of $d\sigma/dt$ on $t$ \cite{rev-LHC}.
These uncertainties are usually connected with the interaction at large distances.

Furthermore, it was shown \cite{BDL,ETPM} that, at small impact parameter
$b$,
the overlapping function grows and may reach a saturation regime
at the LHC, which leads to essential changes in the $t$-dependence of
$\rho$ and of the slope $B$. Hence, the measurements of forward
elastic scattering will give us
information on hadron interactions at
small $b$ and large distances.

\section{Determination of $\sigma_{{\lowercase{\bf tot}}}({\lowercase{\bf s}})$ and $\rho({\lowercase{ \bf s,t}})$}
The experimental analyses of the differential elastic cross section usually rely on the following assumption:
 \begin{equation}
 Im F(s,t) \sim Re F(s,t) \sim e^{ B t/2}.\label{exp}
\end{equation}
Let us note that in the fitting procedure of $d\sigma/dt$, one must vary
at least four parameters, $\sigma_{tot}(s)$, $B(s,t)$, $\rho(s,t)$ and
$n$, the normalisation coefficient reflecting the systematic errors (the
fitted luminosity is $n$ times the true one).
Eq.~(\ref{exp}) works reasonably well at low energies. At high
energies, the elastic amplitude becomes a sum of many different diagrams
which unitarize it and this simple behaviour changes.

In some models \cite{ETPM,DDM}, which describe the differential cross
section in a wide region of $s$ and $t$, the amplitude reaches the Black
Disk Limit (BDL).
This leads to a significant change
of the $t$ dependence of $\rho$ and $B$,
as seen in Fig.~1.a and \cite{BDL}.
In this case, the standard fitting procedure by an exponential
will give bad results.
To see this, we can simulate data at $\sqrt{s}=14 \ $TeV
from the model of ref. \cite{DDM}, and then
 analyse them using an exponential form for the scattering
amplitude.
In Fig.~1.b, we show the result: large differences in $\rho$ and $n$
cannot be detected by a change in $\chi^2$.

As a consequence, one can obtain very wide variations of $\sigma_{tot}$
depending on the input $\rho$, as shown in Fig.~2.a. But the value of $\rho$ at
$\sqrt{s}=14 \ $TeV strongly depends on the model, so it must be included
in the fitting procedure.

\begin{figure}
\includegraphics[height=.2\textheight]{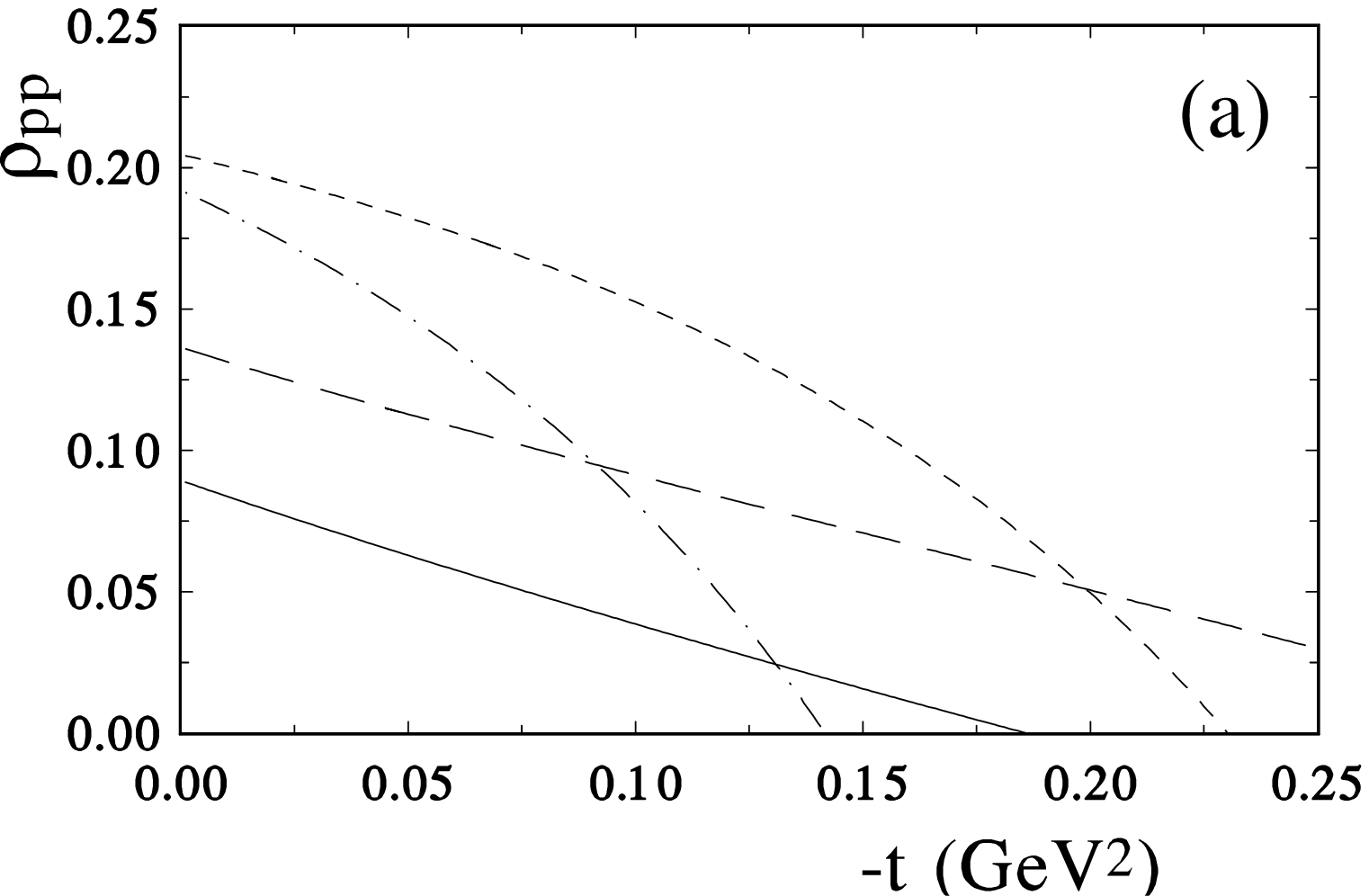}
\includegraphics[height=.2\textheight]{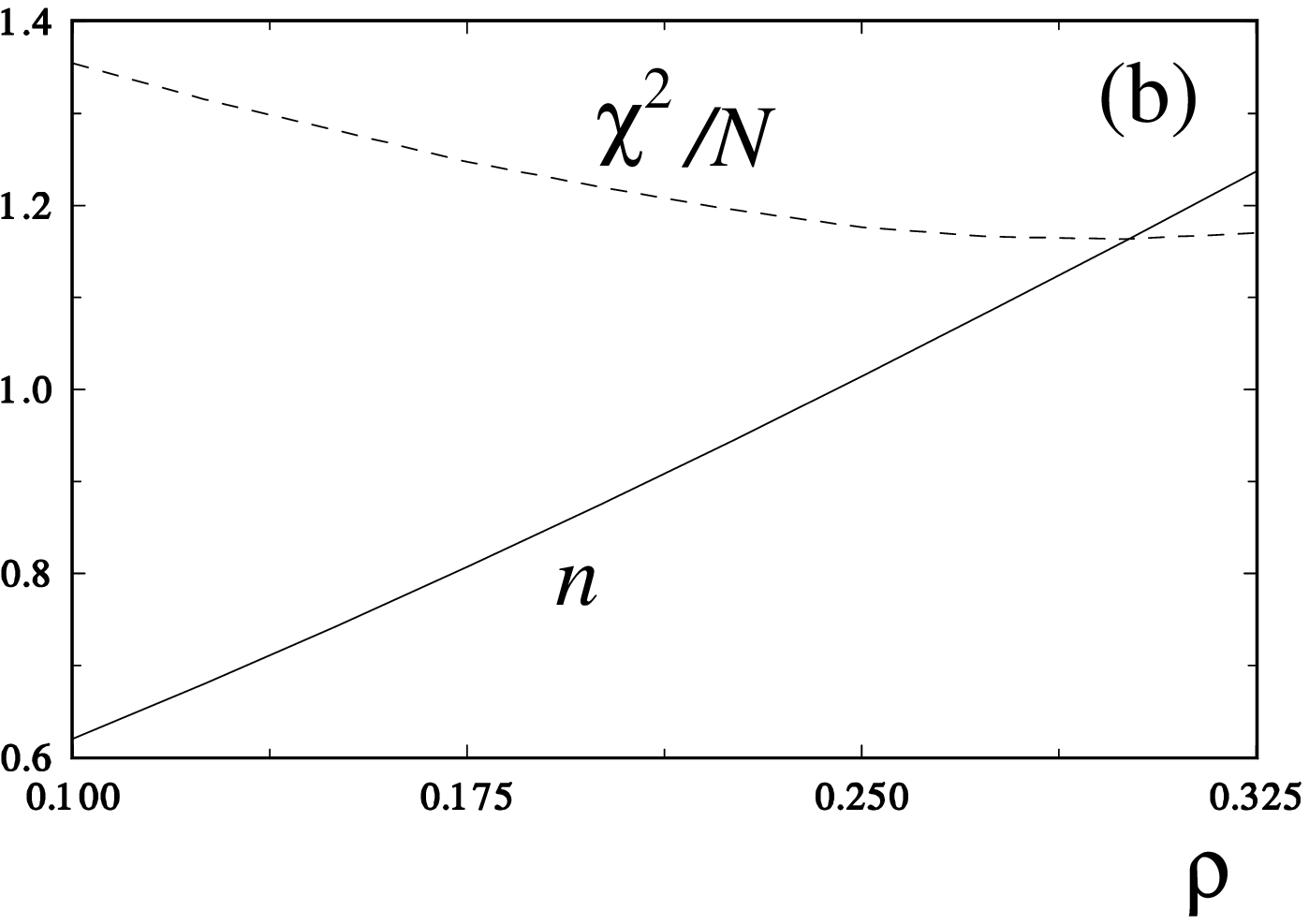}
\caption{  (a) The ratio $\rho(s,t)$ in the Dubna Dynamical Model
\cite{DDM} at $\sqrt{s}=100 \ $GeV
(plain curve), $500 \ $GeV (long dashes), $5 \ $TeV (short dashes) and
$\sqrt{s}=14 \ $TeV
(dash-dotted curve);
( b) The dependence on $\rho$ of $\chi^{2}/N$ and of the
normalisation coefficient $n$ obtained in the fitting procedure
of the simulated data
at $\sqrt{s}=14 \ $TeV.}
\end{figure}

Note also that the slope of the differential cross sections
does not necessarily grow with $s$ \cite{PL94}.  Some models predict a non-vanishing
spin-flip amplitude at high energies. Such a contribution
to the differential cross section will lead to a decrease of the
slope as $t \rightarrow 0$.
Indeed, let us assume that the slope of the spin-flip amplitude,
$F_{sf}/\sqrt{|t|}(t=0)=\tau F_{snf}(t=0)$,
is $\Delta B/2$ larger than the slope $B_0$
of the spin-non-flip amplitude. In
this case, we have for the total slope \cite{Slope}
 \begin{eqnarray}
B(t) = B_{0} - \frac{2 \tau^2 (1+ t  \Delta B )}{m^2+2 \tau^2|t|
\exp(\Delta
B t)}
 e^{\Delta B t},
 \end{eqnarray}
and for the cross section:
  \begin{eqnarray}
\frac{d\sigma}{dt} = \frac{1 }{16 \pi} (1+\rho^{2})\sigma_{tot}^2(1- 2
\tau^2 \frac{t}{m^2} e^{\Delta B_{0} t})
 e^{B_{0}t}.
 \end{eqnarray}
%
\begin{figure}

\includegraphics[width=.45\textwidth]{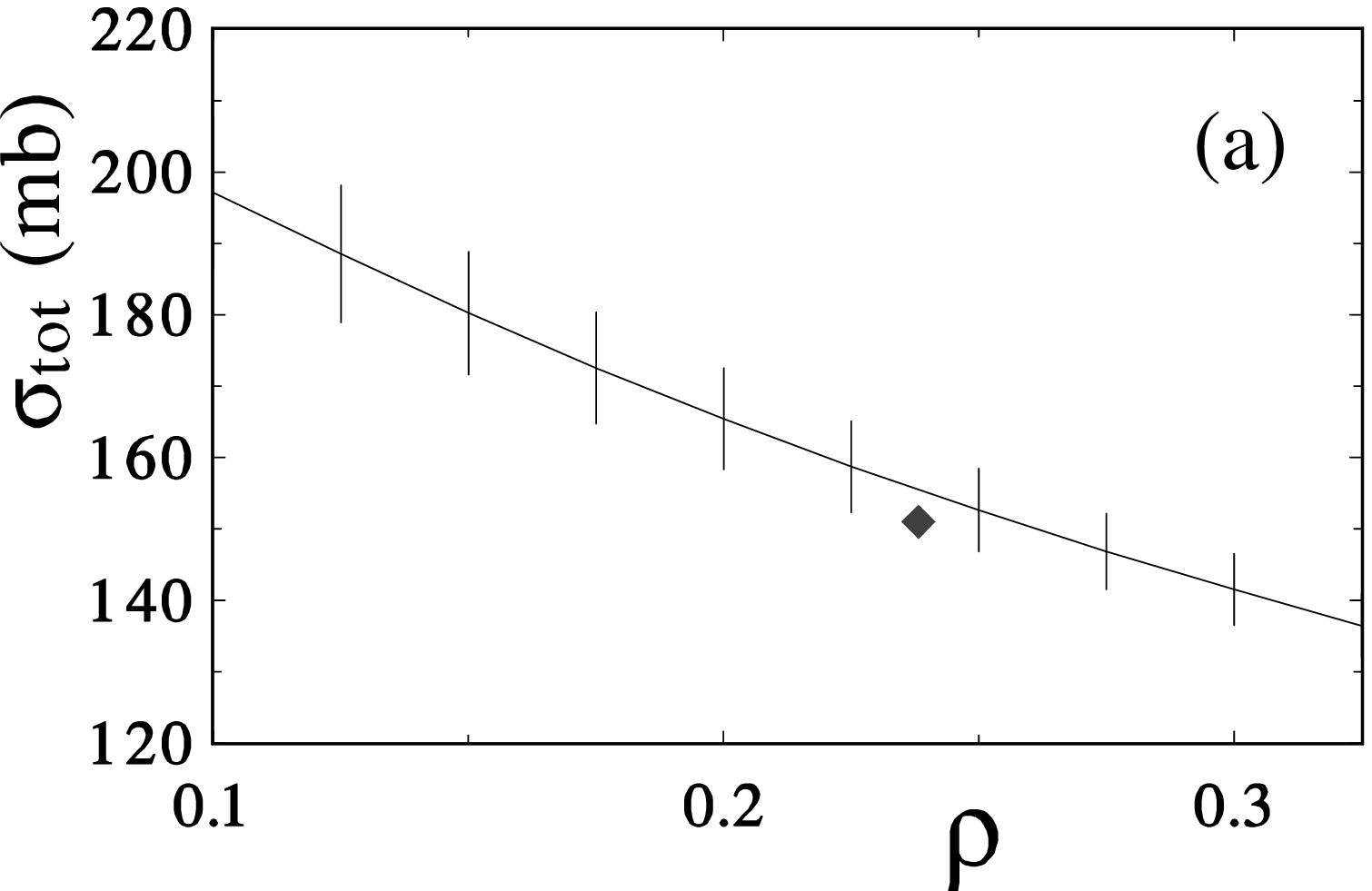}
\includegraphics[width=.33\textwidth]{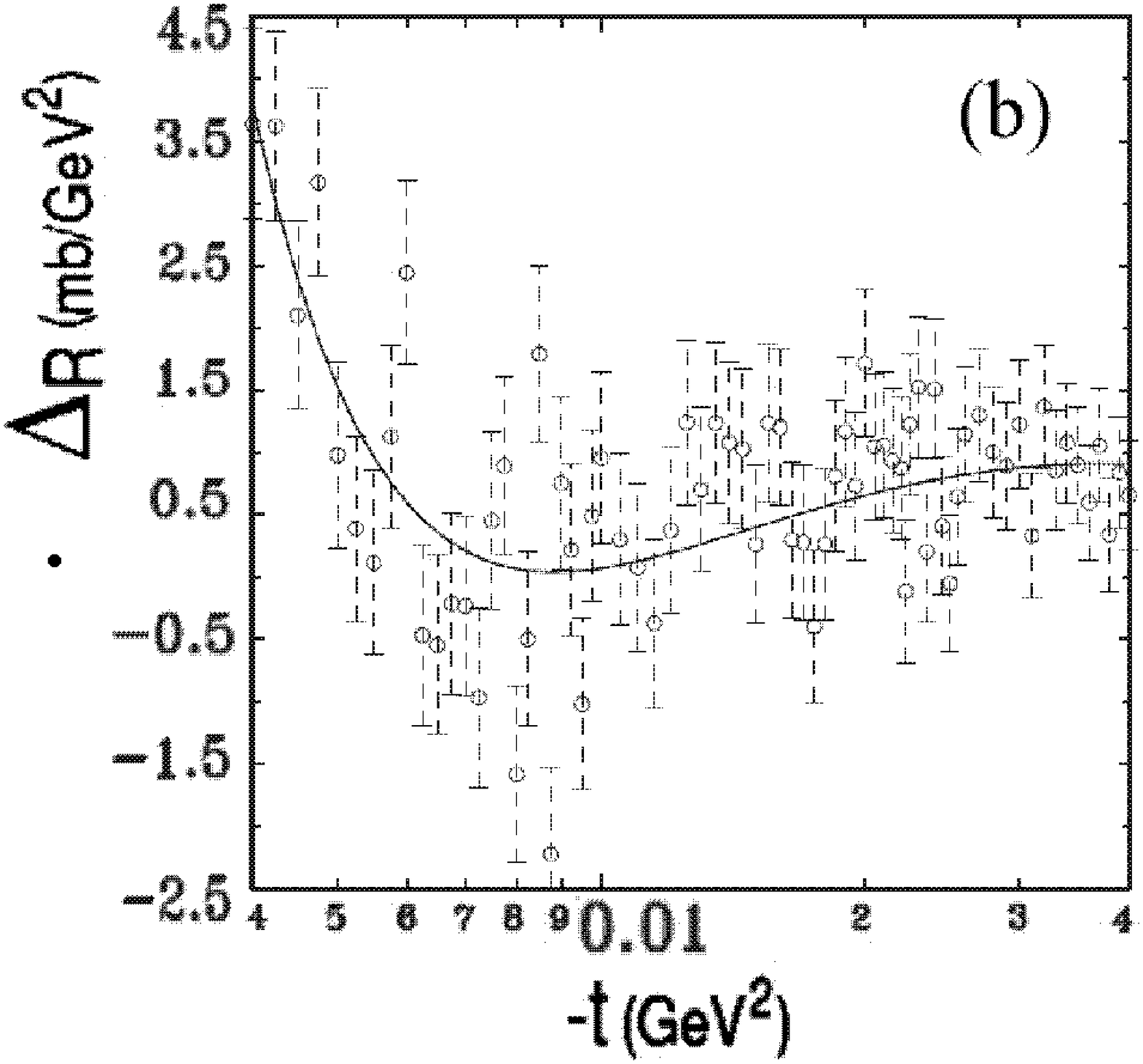}
\caption{(a) $\sigma_{tot}$ obtained from the standard fitting procedure
using an exponential form of the amplitude and a fixed
$\rho$ at $\sqrt{s}=14 \ $TeV with free normalization ($n$);
the diamond gives the input $\sigma_{tot}$ and $\rho$;
(b) $\Delta_R$ obtained  from the simulated data
 at $\sqrt{s}=14 \ $TeV analysed with an exponential form of the scattering amplitude(points with errors); the curve gives the model prediction with $\rho=0.15$.  }
\label{Fig_1}
\end{figure}
\section{Features of the  {\lowercase { \bf t}}-dependence of the real part}
In order to measure reliably $\sigma_{tot}$, we need to find
additional ways to determine $\rho(s,t)$.
One possibility was given in \cite{DelR}, which uses
a particular property  of the proton-proton
scattering amplitude at high energies and small $t$: at a specific value
$t=t_R$, the real part of the Coulombic amplitude cancels the real part
of the hadronic amplitude $Re F_C(t_R) + Re F_h(t_R) = 0$. So, the value
$\Delta_R = (Re F_C + Re F_h)^2 $ will have a minimum at this special
point, and the value of $t_R$ may be possible to
 extract from the experimental data. Note that this is a generic property
that does not rely on a given model. One has in general

\begin{eqnarray}
\Delta_{R}(s,t) &=&  [ Re F_{h}(s,t)+Re F_{C}(s,t)]^2 \label{Del}
\nonumber  \\
&=&[(1/\pi) \  d\sigma^{exp}/ dt
   - (\alpha \varphi F_{C}(t)+Im F_{h}(s,t))^2].
                                               \label{Del2}
\end{eqnarray}
We show in Fig.~2.b what such an analysis might give, for the simulated data at $\sqrt{s} = 14 \ $TeV, fitted
by a simple exponential form with $\rho=0.15$.
\begin{figure}
\includegraphics[height=.2\textheight]{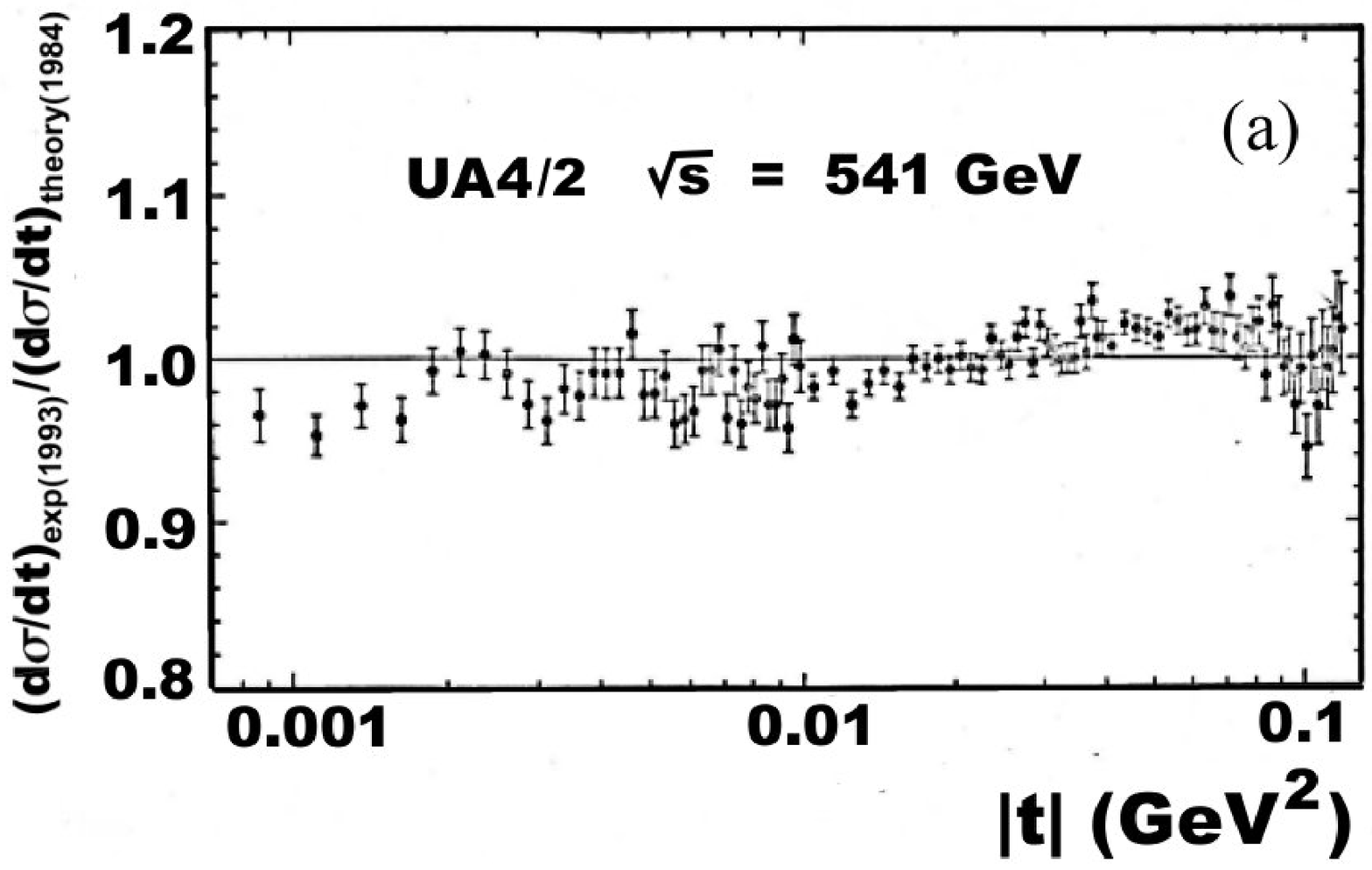}
\includegraphics[height=.18\textheight]{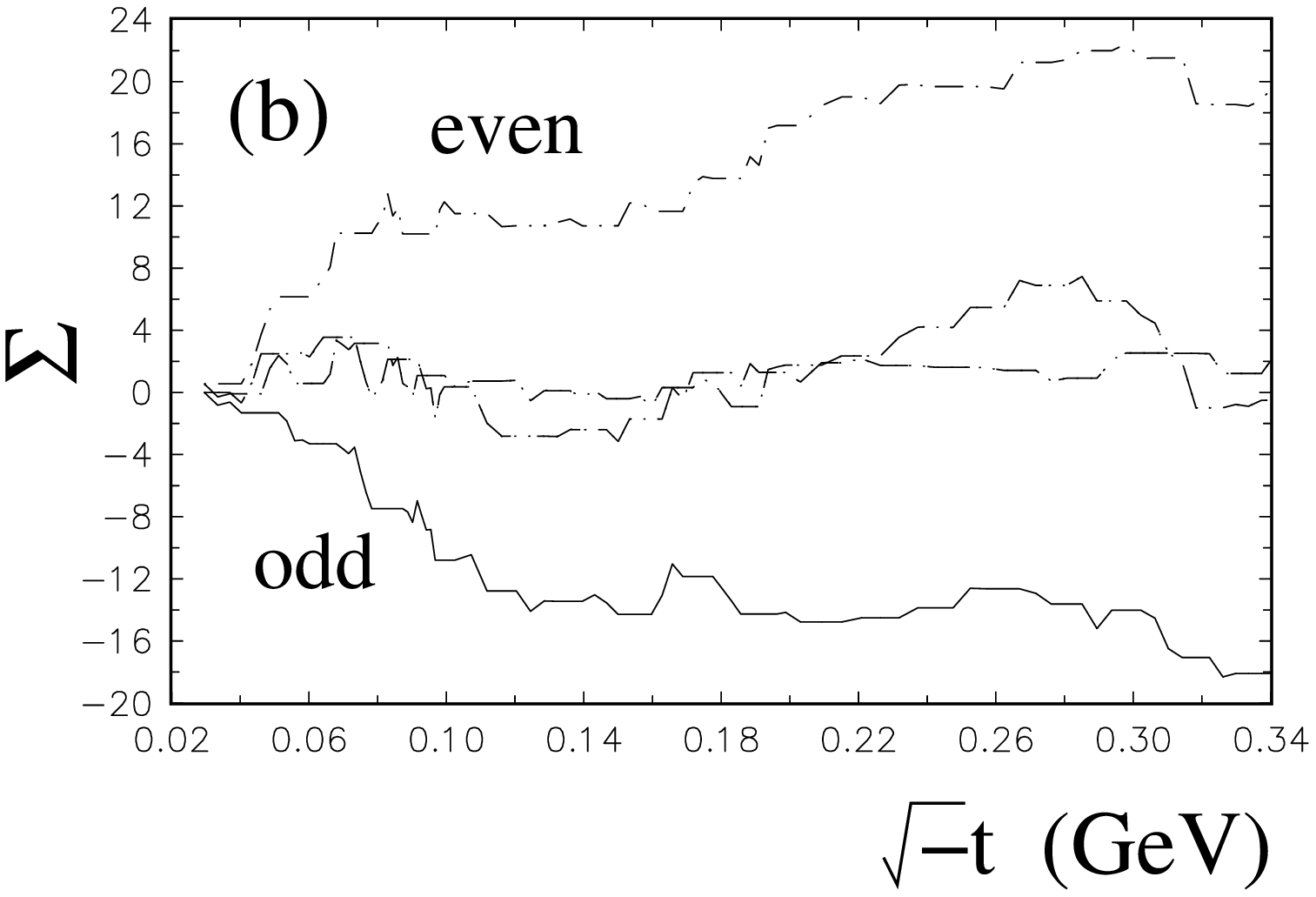}
\caption{ (a) Ratio of the experimental data to the exponential description at $\sqrt{s} = 541 \ $GeV;  (b) The statistical sums (upper and lower curves) of the even and odd intervals coincides with possible oscillations; the middle curves are
the same sums but the beginning of the intervals was shifted by half an interval.
}
\end{figure}

There can be some additional effects which change the form  of
$d\sigma/dt$. For example, there can be some small oscillations in the
scattering amplitude which are predicted by several models  (see
\cite{rev-LHC} for a review ).  One such possible oscillation with period
proportional to $q=\sqrt{|t|}$ was analyzed in \cite{Oscil}. It was shown
that such oscillations may exist in the experimental data obtained at
$\sqrt{s}= 541 \ $GeV, as shown in Fig.~3.a.
However in the standard fitting procedure, if we do not now the form and
$t$-dependence of such oscillations, they are hidden in the statistical
noise.
In \cite{Oscil}, a new method was proposed, based on statistically
independent choices.
The experimental interval in $\sqrt|t|$ is divided into small equal intervals $\delta_i$.
One then calculates sums of the differences between the experimental data and the theoretical curve
normalised by the experimental error, for $i$ odd or even:
\begin{equation}
\Sigma^{odd\ (even)}=\sum_{i\ odd\ (i\ even)}\frac{
d\sigma/dt_i^{data}-d\sigma/dt_i^{theory}}{\sigma_i}
\end{equation}
The corresponding curves are shown
in Fig.~3.b.
  The statistical analysis
has shown  the existence
of oscillations with a period $ \sim \sqrt{|t|}$
at a significance level of 3 $\sigma$.

We can conclude that
some additional research is needed.
It is likely that the BDL regime will be reached at LHC energies.
Its effects will have to be taken into account and one will have to fit
simultaneously all 4 parameters  $n$, $\sigma_{tot}$, $B(s,t)$ and
$\rho(s,t)$. To investigate the non-linear behaviour of the parameters of
the scattering amplitude, one will probably need to develop new methods. \\

\noindent{\bf Acknowledgments:}
 O.V. Selyugin is grateful to the Organizing Committee of the conference for the financial support and the warm hospitality, and also wishes to acknowledge support from FNRS and from the Universit\'e de Li\`ege where part of this work was done.

\end{document}